\documentclass[apj]{emulateapj}

\def\b{b}
\def\c{c}

\usepackage[dvipdfmx]{color}
\usepackage{bm}

\usepackage{aas_macros} 
\usepackage{float}
\usepackage{amsmath,amssymb}
\usepackage{slashbox}
\usepackage{url}
\slugcomment{Published version: February 12, 2014}
\shortauthors{Masuda}
\shorttitle{Very Low-Density planets in the Kepler-51 system
}
\begin{document}
\title{Very Low-Density Planets around Kepler-51 Revealed with Transit Timing Variations and
an Anomaly Similar to a Planet-Planet Eclipse Event}
\author{
Kento Masuda
}
\affil{Department of Physics, The University of Tokyo, Tokyo 113-0033, Japan}
\email{masuda@utap.phys.s.u-tokyo.ac.jp}
\begin{abstract}
We present an analysis of the transit timing variations (TTVs) in the multi-transiting planetary system 
around Kepler-51 (KOI-620). 
This system consists of two confirmed transiting planets, Kepler-51b ($P_{\rm b} = 45.2\,\mathrm{days}$)
and Kepler-51c ($P_{\rm c} = 85.3\,\mathrm{days}$), and one transiting planet candidate KOI-620.02 ($P_{02} = 130.2\,\mathrm{days}$),
which lie close to a $1:2:3$ resonance chain.
Our analysis shows that their TTVs are consistently explained by the three-planet model, and constrains their masses as 
$M_{\rm b} = 2.1_{-0.8}^{+1.5} M_{\oplus}$ (Kepler-51b), 
$M_{\rm c} = 4.0 \pm 0.4 M_{\oplus}$ (Kepler-51c), and 
$M_{02} = 7.6 \pm 1.1 M_{\oplus}$ (KOI-620.02),
thus confirming KOI-620.02 as a planet in this system.
The masses inferred from the TTVs are rather small compared to the planetary radii based on
the stellar density and planet-to-star radius ratios determined from the transit light curves.
Combining these estimates, we find that all three planets in this system have densities among the lowest determined, $\rho_p \lesssim 0.05\,{\rm g\,cm^{-3}}$.
With this feature, the Kepler-51 system serves as another example of low-density compact multi-transiting planetary systems.
We also identify a curious feature in the archived {\it Kepler} light curve during the double transit of Kepler-51b and KOI-620.02,
which could be explained by their overlapping on the stellar disk (a planet-planet eclipse).
If this is really the case, the sky-plane inclination of KOI-620.02's orbit relative to that of Kepler-51b is given by 
$\Delta \Omega = -25.3_{-6.8}^{+6.2}\mathrm{deg}$, implying significant misalignment of their orbital planes. 
This interpretation, however, seems unlikely because such an event that is consistent with all of the observations
is found to be exceedingly rare.
\end{abstract}
\section{Introduction} \label{sec:intro}
Since its launch in 2009, NASA's {\it Kepler} telescope has discovered more than 3000 transiting planetary candidates,
more than one third of which are the members of multi-transiting systems \citep{2013ApJS..204...24B}.
With longer observation spans, the detection limits have been extended to smaller and longer-period planets,
and even a compact solar-system analog has been recently discovered \citep{2014ApJ...781...18C}.
Characterization of such multi-planetary systems (not necessarily transiting) is especially valuable because reproducing the architecture of
multiple planets is theoretically more demanding than for a single planet, and hence the resulting constraints on
the formation theories become much tighter.

Fortunately, multi-transiting systems exhibit various features advantageous for their confirmation and characterization \citep[e.g.,][]{2010arXiv1006.3727R}.
These include transit timing variations \citep[TTVs,][]{2002ApJ...564.1019M, 2005Sci...307.1288H, 2005MNRAS.359..567A}, deviations of transit times from the 
strict periodicity due to the mutual gravitational interaction among the planets.
By modeling the TTVs of several transiting planets consistently,
we can precisely estimate their masses, which are usually inaccessible with the photometric observations alone.
The TTVs become especially prominent for planets in near mean-motion resonances \citep{2005MNRAS.359..567A},
which turned out to amount to several percent of the {\it Kepler} sample \citep{2013ApJS..208...16M, 2013arXiv1308.3751X}.
Indeed, planetary masses obtained from TTVs are all the more valuable for systems for which radial velocities are difficult to obtain,
as are the cases for most of the {\it Kepler} target stars.
	
A wealth of the {\it Kepler} data has also revealed a new phenomenon called a planet-planet eclipse (PPE), which was
first and only observed in the Kepler-89 (KOI-94) system \citep{2012ApJ...759L..36H, 2013ApJ...778..185M}. 
In this event, two planets transit the host star simultaneously and even overlap with each other on the stellar disk.
In addition to its rareness and astronomical interest, this phenomenon tightly constrains the relative angular momentum of the
two planets involved, which could give some clues to unveil the history of their orbital evolutions through 
the presence or absence of dynamical planet-planet interaction.
	
This paper focuses on the Kepler-51 (KOI-620) system, one of the multi-transiting systems found by {\it Kepler}.
This system hosts three transiting planet candidates, two of which were confirmed by \citet{2013MNRAS.428.1077S}.
They made sure that these two planets, Kepler-51b (KOI-620.01) and Kepler-51c (KOI-620.03),
are revolving around the same star by confirming that their TTVs are anti-correlated, 
and showed that they are indeed planetary by giving mass upper limits based on the long-term stability of the system.
In this paper, we perform a numerical analysis of their TTVs 
to more fully characterize the system 
and to confirm that the mass of KOI-620.02 is also in the planetary range.
We also discuss a very intriguing light curve of Kepler-51 recently made public on the 
Space Telescope Science Institute (STScI) MAST archive, 
which shows a feature similar to a PPE event.
\section{Stellar and Planet Properties} \label{sec:parameters}
We adopt the stellar properties in Table \ref{par_star} taken from the NASA Exoplanet Archive.\footnote{http://exoplanetarchive.ipac.caltech.edu/}
As an initial guess for the limb-darkening coefficients for the quadratic law, we adopt $(u_1, u_2) = (0.36, 0.28)$,
the values for $(T_{\rm eff}, \log g, Z, \xi) = (6000\,{\rm K}, 4.5\,{\rm dex}, 0.0, 0.0\,{\rm km\,s^{-1}})$
in the grid of \citet{2011A&A...529A..75C}.
Linear ephemerides and transit parameters are retrieved from the MAST archive (Table \ref{ppar_prior})
as a starting point for the iterative determination of these parameters in Section \ref{sec:transit}.
\begin{table}
	\begin{center}
	\caption{Stellar properties of Kepler-51 (KOI-620)}
	\label{par_star}
	\begin{tabular}{c@{\hspace{3cm}}c}
	\tableline \tableline
	Parameter		&		Value\\
	\tableline
	$K_{\rm p}$						&	$14.669$\\
	$T_{\mathrm{eff}}\,(\mathrm{K})$		&	$6018 \pm 107$\\
	$\log g \,(\mathrm{dex})$			&	$4.510 \pm 0.300$\\
	$M_{\star} (M_{\odot})$				&	$1.04 \pm 0.12$\\
	$R_{\star} (R_{\odot})$				&	$0.940 \pm 0.500$\\
	Age\,(Gyr)						&	$0.3 \pm 2.3$\\
	\tableline 
	\end{tabular}
	\end{center}
\end{table}

\begin{table*}
	\begin{center}
	\caption{Properties of the Kepler-51 system determined by other authors}
	\label{ppar_prior}
	\begin{tabular}{cccc}
	\tableline \tableline
	 Parameter			&	Kepler-51b			&	Kepler-51c				&	KOI-620.02\\				
	\tableline
	\multicolumn{4}{c}{\it Transit parameters determined by the Kepler team\tablenotemark{a}}\\
	\tableline
	$t_0$ (BJD - 2454833) & $159.10435 \pm 0.00062$ 	& $295.321 \pm 0.002$ 			& $212.02345 \pm 0.00062$\\
	$P$ (days)		& $45.155503 \pm 0.000072$ 	& $85.31287\pm 0.00096$		& $130.1831 \pm 0.00033$\\
	$a / R_{\star}$ 	& $63.880 \pm 0.640$	 		& $97.630 \pm 0.970$ 			& $129.400 \pm 1.300$\\
	$R_p / R_{\star}$ 	& $0.07074 \pm 0.00020$ 		& $0.0573 \pm 0.0081$ 			& $0.0972 \pm 0.00024$\\
	$b$ 				& $0.030 \pm 0.020$ 			& $0.972 \pm 0.028$ 			& $0.061 \pm 0.010$\\
	$\rho_{\star} ({\rm g\,cm^{-3}})$	&	\multicolumn{3}{c}{$2.42 \pm 0.07$}\\
	\tableline
	\multicolumn{4}{c}{{\it  Mass limit from the stability analysis} \citep{2013MNRAS.428.1077S}}\\
	\tableline
	Maximum mass ($M_{\rm J}$)	& $3.23$				& $2.60$				& -\\	
	\tableline
	\end{tabular}
	\tablenotetext{1}{Data from the MAST archive \url{http://archive.stsci.edu/kepler/}.}
	\end{center}
\end{table*}

\section{Analysis of the Transit Light Curves} \label{sec:transit}
\subsection{Data Processing} \label{sec:dataprocess}
We analyze the short-cadence ($\sim 1$ minute)
Pre-search Data Conditioned Simple Aperture Photometry (PDCSAP)
fluxes from Quarters $12$ to $16$
as well as the long-cadence ($\sim 30$ minutes) fluxes from Quarters $1$ to $11$,
for which short-cadence data are not available.
We first extract data points within $\pm 1$ day of every transit, and
iteratively fit the points outside the transit 
with a third-order polynomial until all the out-of-transit outliers exceeding $5\sigma$ are excluded.
Then we divide all the points in the chunk by the best-fit polynomial to give a detrended and normalized transit light curve.
Also excluding the transits that are not fully observed, we obtain $30$, $11$, and $10$ transits for Kepler-51b, Kepler-51c, and KOI-620.02, respectively.
We note that the Kepler-51b's transit around ${\rm BJD} = 2456346.8$ occurred simultaneously with that of KOI-620.02 (double transit).
Since this particular transit shows a possible sign of a PPE \citep{2012ApJ...759L..36H, 2013ApJ...778..185M}, 
we will discuss it in more detail in Section \ref{sec:PPE}.

\subsection{Transit Times and Transit Parameters}
From the transit light curves obtained in Section \ref{sec:dataprocess}, we determine the transit times and transit parameters of 
the three planets by iterative fit using a Markov chain Monte Carlo (MCMC) algorithm. We repeat the following two steps:
(i) We fit each transit for the time of the transit center $t_c$ using the light curve model by \citet{2009ApJ...690....1O}.
Here we assume $e=0$ and fix the values of $P$, $R_p/R_{\star}$, $b$ (of each planet), $u_1$, $u_2$, and $\rho_{\star}$.
From the series of transit times, period $P$ and time of a transit center $t_0$ are extracted by linear regression.
(ii) Using the transit times obtained in step (i), we phase fold the transits of each planet and fit the three phase curves simultaneously 
for $R_p/R_{\star}$, $b$ (of each planet), $u_1$, $u_2$, and $\rho_{\star}$.
Here the values of $P$ are fixed at those in step (i) and $e=0$ is assumed for all the planets.

\begin{figure}
	\begin{center} 
	\includegraphics[width=9.3cm,clip]{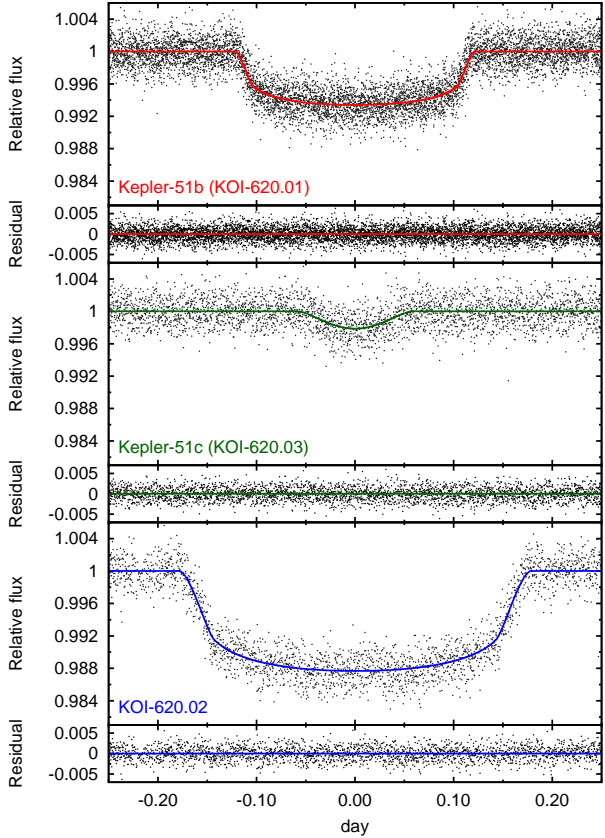}
	\caption{Phase-folded transit light curves of Kepler-51b (top), Kepler-51c (middle), and KOI-620.02 (bottom).
	Black dots are the observed fluxes and colored solid lines show the best-fit models.}
	\label{phase}
	\end{center}
\end{figure}

Starting from the values in Table \ref{ppar_prior} (and in Section \ref{sec:parameters} for $u_1$ and $u_2$),
all the parameters converge sufficiently well after five iterations.
The resulting transit parameters, ephemerides, and transit times are summarized in Tables \ref{ppar_5th} to \ref{tc_02}.
The quoted best-fit parameters ($a/R_\star$ to $\rho_\star$ and $t_c$) denote the median values of their posteriors,
and uncertainties exclude $15.87\%$ of values at upper and lower extremes.
The corresponding best-fit transit models with the phase-folded transits are shown in Figure \ref{phase}.
As reported in previous analyses \citep{2013MNRAS.428.1077S, 2013ApJS..208...16M}, 
we find significant TTVs for all three planets as shown in Figure \ref{ttv_fit}.
Note that the TTV amplitude of KOI-620.02 in our analysis is 
about twice as large as that first reported by \citet{2013ApJS..208...16M}, who analyzed the first twelve quarters of {\it Kepler} data.

Several comments should be added to our revised values of transit parameters in comparison to those in Table \ref{ppar_prior}.
With the longer baselines, we refine the orbital periods of the three planets with better precision than the previous values.
We also find the larger values for $R_p/R_{\star}$,  albeit with relatively large uncertainties.
This is because the slight variations of transit depths we identify in the archived light curves, probably due to the
star-spot activities (see also the discussion in Section \ref{sec:PPE}).
In fact, our analysis completely neglects such spot effects, and so the values of $R_p/R_{\star}$ we determined may be overestimated.
The constraint on $R_p/R_{\star}$ is especially poor for Kepler-51c, whose grazing transit causes
the strong correlation between its planetary radius and impact parameter.
The values of impact parameters we determine are marginally consistent with those in Table \ref{ppar_prior}, 
corresponding to the slightly different value of stellar density;
these parameters would be determined more precisely with spectroscopic constraints on the stellar mass and radius.

In addition, we find that the difference between the impact parameters of Kepler-51b and KOI-620.02 is tightly constrained 
due to the strong correlation, in spite of their relatively large uncertainties:
the MCMC posteriors for the two parameters yield 
$b\,({\rm KOI\mathchar`-620.02}) - b\,({\rm Kepler\mathchar`-51b}) = -0.001 \pm 0.01$.
Since this difference is closely related to the minimum separation during a simultaneous transit of the two planets,
it has an important role in assessing the occurrence of the PPE, as will be discussed in Section \ref{sec:PPE}.

\begin{table*}
	\begin{center}
	\caption{Revised transit parameters obtained from the phase-folded transit light curves} 
	\label{ppar_5th}
	\begin{tabular}{cccc}
	\tableline \tableline
	 Parameter			&		Kepler-51b				&	Kepler-51c					&	KOI-620.02\\
	\tableline
	$t_0$ $({\rm BJD} - 2454833)$ 	& $159.10653 \pm 0.00033$ 	& $295.3131 \pm 0.0018$ 			& $212.03246 \pm 0.00039$\\
	$P$ (days)				& $45.155314 \pm 0.000019$ 	& $85.31644 \pm 0.00022$		& $130.178058 \pm 0.000071$\\
	$a / R_{\star}$ 			& $61.5_{-1.2}^{+1.5}$				& $94.1_{-1.9}^{+2.2}$ 	  			& $124.7_{-2.5}^{+3.0}$		\\
	$R_p / R_{\star}$ 		& $0.07414_{-0.00061}^{+0.00059}$	& $0.094_{-0.017}^{+0.028}$ 			& $0.10141_{-0.00085}^{+0.00084}$\\
	$b$ 					& $0.251_{-0.138}^{+0.073}$			& $1.017_{-0.023}^{+0.034}$			& $0.250_{-0.141}^{+0.075}$		\\
	$u_1$ 				& \multicolumn{3}{c}{$0.375_{-0.036}^{+0.040}$} 		\\
	$u_2$ 				& \multicolumn{3}{c}{$0.311_{-0.087}^{+0.083}$} 			\\
	$\rho_{\star}\,({\rm g\,cm^{-3}})$	& \multicolumn{3}{c}{$2.16_{-0.13}^{+0.15}$} 	\\
	$\chi^2/\mathrm{d.o.f}$	&  \multicolumn{3}{c}{$12681/12417$} 							\\
	\tableline
	\end{tabular}
	\end{center}
\end{table*}
\begin{table}
	\centering
	\caption{Transit times of Kepler-51$\rm{\b}$ (KOI-620.01)}
	\label{tc_01}
	\scriptsize
	\begin{tabular}{cccccc}
	\tableline \tableline
	Transit 	&	$t_c$ 	&	$1\sigma_{\rm lower}$	&	$1\sigma_{\rm upper}$	&	$\chi^2/{\rm d.o.f}$	&	$	O - C	$\\number&$({\rm BJD} - 2454833)$ & & & & (days)\\
	\tableline
	0	&	159.10975 	&	0.00072 	&	0.00072 	&	2.14 	&	$	0.00323 	$	\\
	1	&	204.26437 	&	0.00078 	&	0.00076 	&	1.86 	&	$	0.00253 	$	\\
	2	&	249.41453 	&	0.00120 	&	0.00152 	&	3.24 	&	$	-0.00262 	$	\\
	3	&	294.57446 	&	0.00251 	&	0.00159 	&	2.12 	&	$	0.00199 	$	\\
	4	&	339.72399 	&	0.00083 	&	0.00088 	&	2.32 	&	$	-0.00379 	$	\\
	5	&	384.87799 	&	0.00078 	&	0.00079 	&	4.04 	&	$	-0.00510 	$	\\
	6	&	430.03405 	&	0.00076 	&	0.00076 	&	1.78 	&	$	-0.00436 	$	\\
	8	&	520.34240 	&	0.00151 	&	0.00168 	&	0.80 	&	$	-0.00663 	$	\\
	9	&	565.49926 	&	0.00106 	&	0.00148 	&	3.29 	&	$	-0.00509 	$	\\
	10	&	610.65682 	&	0.00087 	&	0.00095 	&	1.00 	&	$	-0.00285 	$	\\
	11	&	655.81302 	&	0.00080 	&	0.00084 	&	1.38 	&	$	-0.00196 	$	\\
	12	&	700.97595 	&	0.00204 	&	0.00156 	&	2.19 	&	$	0.00566 	$	\\
	13	&	746.12646 	&	0.00082 	&	0.00086 	&	1.10 	&	$	0.00085 	$	\\
	14	&	791.28654 	&	0.00102 	&	0.00129 	&	1.79 	&	$	0.00562 	$	\\
	15	&	836.43982 	&	0.00074 	&	0.00074 	&	2.24 	&	$	0.00358 	$	\\
	16	&	881.59882 	&	0.00072 	&	0.00071 	&	0.91 	&	$	0.00727 	$	\\
	17	&	926.75475 	&	0.00083 	&	0.00078 	&	1.42 	&	$	0.00789 	$	\\
	18	&	971.90566 	&	0.00181 	&	0.00262 	&	1.95 	&	$	0.00348 	$	\\
	19	&	1017.05878 	&	0.00083 	&	0.00088 	&	1.62 	&	$	0.00129 	$	\\
	20	&	1062.21217 	&	0.00075 	&	0.00075 	&	2.50 	&	$	-0.00064 	$	\\
	21	&	1107.36887 	&	0.00095 	&	0.00097 	&	0.94 	&	$	0.00075 	$	\\
	22	&	1152.52090 	&	0.00088 	&	0.00088 	&	0.96 	&	$	-0.00253 	$	\\
	23	&	1197.67687 	&	0.00097 	&	0.00097 	&	0.87 	&	$	-0.00188 	$	\\
	24	&	1242.83059 	&	0.00087 	&	0.00087 	&	0.99 	&	$	-0.00347 	$	\\
	25	&	1287.98482 	&	0.00086 	&	0.00088 	&	0.92 	&	$	-0.00456 	$	\\
	26	&	1333.14289 	&	0.00091 	&	0.00090 	&	0.95 	&	$	-0.00179 	$	\\
	27	&	1378.29779 	&	0.00088 	&	0.00088 	&	0.86 	&	$	-0.00220 	$	\\
	28	&	1423.45442 	&	0.00091 	&	0.00090 	&	1.00 	&	$	-0.00089 	$	\\
	29	&	1468.61324 	&	0.00089 	&	0.00089 	&	0.97 	&	$	0.00261 	$	\\
	\tableline 
	\end{tabular}
\end{table}
\begin{table}
	\centering
	\caption{Transit times of Kepler-51$\rm{\c}$ (KOI-620.03)}
	\label{tc_03}
	\scriptsize
	\begin{tabular}{cccccc}
	\tableline \tableline
	Transit 	&	$t_c$ 	&	$1\sigma_{\rm lower}$	&	$1\sigma_{\rm upper}$	&	$\chi^2/{\rm d.o.f}$	&	$	O - C$\\number	&$({\rm BJD} - 2454833)$ & & & & (days)\\
	\tableline
	0	&	295.31257 	&	0.00378 	&	0.00384 	&	0.98 	&	$	-0.00057 	$	\\
	1	&	380.64295 	&	0.00358 	&	0.00354 	&	0.97 	&	$	0.01337 	$	\\
	2	&	465.95289 	&	0.00287 	&	0.00283 	&	1.41 	&	$	0.00687 	$	\\
	3	&	551.26161 	&	0.00319 	&	0.00304 	&	0.99 	&	$	-0.00086 	$	\\
	4	&	636.56677 	&	0.00324 	&	0.00325 	&	2.04 	&	$	-0.01214 	$	\\
	7	&	892.51469 	&	0.00384 	&	0.00393 	&	1.90 	&	$	-0.01355 	$	\\
	8	&	977.84149 	&	0.00360 	&	0.00364 	&	1.16 	&	$	-0.00319 	$	\\
	10	&	1148.45861 	&	0.00327 	&	0.00327 	&	1.00 	&	$	-0.01896 	$	\\
	11	&	1233.80785 	&	0.00322 	&	0.00324 	&	0.89 	&	$	0.01385 	$	\\
	12	&	1319.11072 	&	0.00331 	&	0.00342 	&	0.95 	&	$	0.00027 	$	\\
	14	&	1489.75414 	&	0.00337 	&	0.00340 	&	0.88 	&	$	0.01080 	$	\\
	\tableline 
	\end{tabular}
\end{table}
\begin{table}
	\centering
	\caption{Transit times of KOI-620.02}
	\label{tc_02}
	\scriptsize
	\begin{tabular}{cccccc}
	\tableline \tableline
	Transit &	$t_c$	& $1\sigma_{\rm lower}$	&	$1\sigma_{\rm upper}$	&	$\chi^2/{\rm d.o.f}$	&	$O - C$ \\number
	&$({\rm BJD} - 2454833)$ & & & & (days)\\
	\tableline
	0	&	212.02417 	&	0.00066 	&	0.00066 	&	2.67 	&	$	-0.00829 	$	\\
	1	&	342.20715 	&	0.00063 	&	0.00062 	&	2.28 	&	$	-0.00337 	$	\\
	2	&	472.39116 	&	0.00064 	&	0.00064 	&	2.08 	&	$	0.00258 	$	\\
	3	&	602.57341 	&	0.00063 	&	0.00063 	&	2.17 	&	$	0.00678 	$	\\
	5	&	862.93196 	&	0.00076 	&	0.00070 	&	3.88 	&	$	0.00921 	$	\\
	6	&	993.10424 	&	0.00064 	&	0.00065 	&	2.35 	&	$	0.00343 	$	\\
	7	&	1123.28307 	&	0.00065 	&	0.00066 	&	1.12 	&	$	0.00420 	$	\\
	8	&	1253.44963 	&	0.00062 	&	0.00063 	&	0.89 	&	$	-0.00730 	$	\\
	9	&	1383.62994 	&	0.00064 	&	0.00064 	&	0.99 	&	$	-0.00505 	$	\\
	\tableline
	\end{tabular}
\end{table}
\section{TTV Analysis} \label{sec:TTV}
In this section, we perform a numerical analysis of the TTVs for the planetary parameters (especially their masses)
to confirm KOI-620.02 as a planet and to more fully characterize the system.
For simplicity, we assume the coplanar orbits for the three planets and fix the stellar mass at $M_{\star} = 1.04 M_{\odot}$.
We make no attempt to model transit parameters other than the transit time.

We define transit centers as the minima of the star-planet distance in the plane of the sky ($D$), and calculate the simulated transit times in the following way.
We integrate the planetary orbits using the fourth-order Hermite scheme with the shared time step \citep{2004PASJ...56..861K}.
From the position and velocity of each planet,
we calculate the time derivative of $D$
and search for its root applying the Newton-Raphson method \citep{2010arXiv1006.3834F}.
All the simulations presented in this section are performed between $\mathrm{BJD} = 2454980$ and $\mathrm{BJD} = 2456345$,
beginning at the same epoch $T_0 (\mathrm{BJD}) = 2455720$ (close to the center of the observation time).

We fit the three planets' TTVs simultaneously
for the mass $M_p$, transit time closest to the epoch $T_c$, orbital period $P$, eccentricity $e$, and argument of periastron $\omega$
(measured from the sky plane)
of each planet.
Here $P$, $e$, and $\omega$ are the osculating orbital elements defined at the epoch $T_0$.
Since we assume the coplanar orbits, we fix the initial values of the orbital inclinations $i = \pi/2$ and longitudes of the ascending nodes $\Omega = 0$.
Chi-squares are computed from the simulated transit times as 
\begin{align}
	\chi^2 =  \sum_{j:\,{\rm planets}}
			\sum_{\substack{i:\,{\rm observed}\\{\rm transits}}}
			\left[ \frac{t_{c,j}(i) - t^{\rm sim}_{c,j}(i)}{\sigma_j (i)} \right]^2,
\end{align}
where $t^{\rm sim}_{c,j}(i)$, $t_{c,j}(i)$, and $\sigma_j (i)$ are the simulated central time, observed central time, and uncertainty
of the $i$th transit of planet $j$, respectively.
For simplicity, we adopt averages of $1\sigma$ upper and lower limits of transit times as $\sigma_j (i)$.\footnote{
Even taking into account the asymmetry in the posteriors when calculating $\chi^2$, the resulting parameters 
are consistent with those given here, but the value of $\chi^2$ is slightly reduced.}

\begin{figure}
	\begin{center}
	\includegraphics[width=9.5cm,clip]{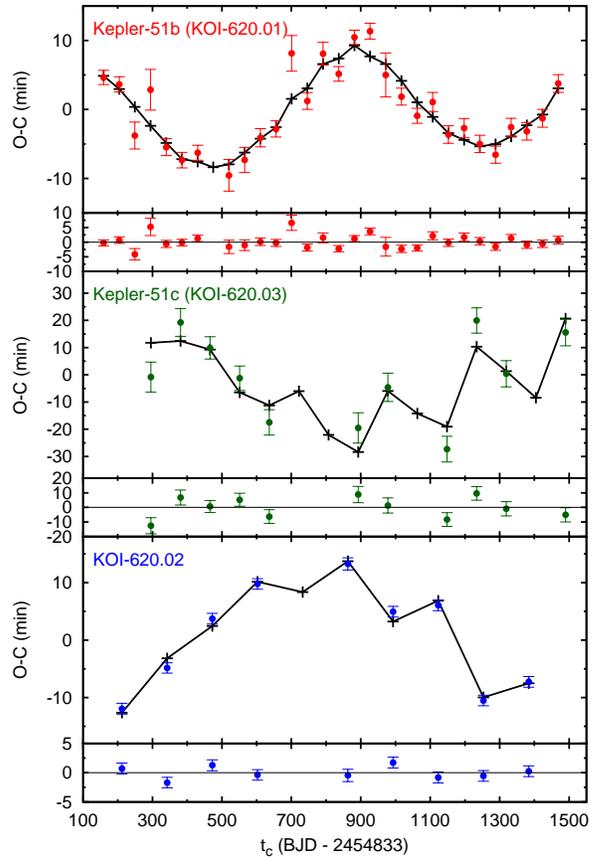}
	\caption{Best-fit numerical models (black solid lines) for the observed TTVs (colored points with error bars).
	Here we adopt the parameters that correspond to the $\chi^2$ minimum, 
	which are slightly different from the median values listed in Table \ref{ttv_result}.}
	\label{ttv_fit}
	\end{center}
\end{figure}
\begin{table*}
	\begin{center}
	\caption{Best-fit parameters obtained from TTVs}
	\label{ttv_result}
	\begin{tabular}{cccc}
	\tableline \tableline
	 Parameter				&		Kepler-51b					&	Kepler-51c					&	KOI-620.02\\
	\tableline													
	$M_p$ ($M_{\oplus}$)		& $2.1_{-0.8}^{+1.5}$			    		& $4.0 \pm 0.4$					& $7.6 \pm 1.1$\\	
	$T_c$ (${\rm BJD}-2454833$)& $881.5977 \pm 0.0004$	 			& $892.509 \pm 0.003$				& $862.9323 \pm 0.0004$\\		
	$P$ (days)				& $45.1540 \pm 0.0002$				& $85.312_{-0.002}^{+0.003}$		& $130.194_{-0.002}^{+0.005}$\\
	$e \cos \omega$ 			& $-0.016 \pm 0.006$					& $0.010_{-0.008}^{+0.013}$ 		& $0.005_{-0.006}^{+0.011}$\\				
	$e \sin \omega$ 			& $-0.04 \pm 0.01$					& $-0.009_{-0.013}^{+0.009}$		& $-0.006_{-0.010}^{+0.008}$\\				
	$\chi^2$					& $51$ (29 transits)					& $21$ (11 transits)				& $11$ (9 transits)	\\		
	$\chi^2$/d.o.f				& \multicolumn{3}{c}{$83/34$}	\\	
	\tableline
	\end{tabular}
        \tablecomments{$P$ in this table is one of the osculating orbital elements at the simulation epoch $T_0$, and so its value is different from the average period obtained from the transits in Table \ref{ppar_5th}.}
	\end{center}
\end{table*}

We first use the downhill simplex method by Nelder and Mead \citep{1992nrca.book.....P} to find the minimum in the above $\chi^2$,
and then perform an MCMC analysis \citep{2005AJ....129.1706F, 2006ApJ...642..505F} around this minimum.
The median values of the MCMC posteriors, their $1\sigma$ uncertainties, and minimum value of $\chi^2$ are shown in Table \ref{ttv_result}, 
and the corresponding best-fit simulated TTVs are plotted in Figure \ref{ttv_fit}.
We perform the same procedures also floating $M_{\star}$ with the Gaussian prior based on $M_{\star} = 1.04 \pm 0.12 \,M_{\odot}$,
and obtain the consistent results with no better constraint on $M_{\star}$.
Analysis taking account of the apparent non-coplanarity of Kepler-51c ($i \sim 89.4^\circ$) does not alter the result, either.

In Figure \ref{ttv_fit}, the sinusoidal TTVs of Kepler-51b and KOI-620.02 are well explained by their proximities to $2:1$ and $3:2$ resonances,
respectively, with Kepler-51c:
the periods of these two planets' TTVs inferred from the observed data ($\sim 770\,{\rm days}$ and $\sim 2500\,{\rm days}$) are in well agreement with
the ``super-period" $P^j = 1/|j/P_{\rm outer} - (j-1)/P_{\rm inner}|$ for a $j:j-1$ resonance defined by \citet{2012ApJ...761..122L}.
In addition, the best-fit masses of all three planets, including the one of KOI-620.02, fall into the planetary regime.
These facts strongly indicate that KOI-620.02 is a planet belonging to the same system as the other two.

Remarkably, the best-fit masses of all three planets are less than that of Neptune in spite of their relatively large values of $R_p / R_{\star}$.
At least for the mass of Kepler-51c, our value is also supported by another study:
\citet{2013arXiv1310.7942H} analyzed the {\it Kepler} data through Quarter 12 using analytic TTV formulae derived by \citet{2012ApJ...761..122L}.
In their formulae, the TTV amplitudes of a pair of coplanar planets in near a $j:j-1$ resonance are analytically given as functions of their masses, eccentricities, 
and orbital phases.
Since the orbital phases are already constrained from transit observations, the formulae allow us to constrain the planets' masses and eccentricities
in a degenerate way.
Assuming $e=0$, they obtain two estimates for the mass of Kepler-51c, 
one using the inner pair ($9.7\,M_\oplus$) and one using the outer pair ($3.1\,M_\oplus$).
Although the non-zero eccentricities easily alter these estimates by a factor of a few \citep{2012ApJ...761..122L}, 
these values are consistent with the mass of Kepler-51c that we obtain here.

Using $\rho_{\star} = 2.16_{-0.13}^{+0.15}\,{\rm g\,cm^{-3}}$ obtained from the transit light curves (Table \ref{ppar_5th})
and $M_{\star} = 1.04 \pm 0.12 M_{\odot}$, we obtain $R_{\star} = 0.88 \pm 0.04 \,R_{\odot}$, 
which is consistent with the value in Table \ref{par_star}.
Note that $e=0$ is assumed in determining the value of $\rho_{\star}$, 
though the correction is of order $e \sin \omega$ and smaller than the uncertainty of $\rho_{\star}$ in Table \ref{ppar_5th}, 
according to the TTV results in Table \ref{ttv_result}.
This value of $R_{\star}$, along with the values of $R_p/R_{\star}$ in Table \ref{ppar_5th} and $M_p$ in Table \ref{ttv_result}, gives the
radius and density of each planet listed in Table \ref{planet}.
In addition to the fact that they have relatively large uncertainties, the planetary radii could be slightly overestimated due to the star spots,
as discussed in Section \ref{sec:transit}.
Nevertheless, these results
show that the bulk densities of all the planets in the Kepler-51 system 
are arguably among the lowest of the known planets,
falling below that of the recently discovered sub-Saturn radius planet KOI-152d \citep{2013arXiv1310.2642J}.

While their densities are much lower than those of the planets in the solar system, 
it is possible to form such planets with sufficiently rich gas envelopes.
\citet{2013arXiv1311.0329L} calculated the radii of low-mass ($1\mathchar`-20 M_{\oplus}$) planets for various values of 
envelope fraction ($0.01\mathchar`-60\,\%$), incident flux ($0.1\mathchar`-1000 F_\oplus$), and age ($10\,{\rm Myr}\mathchar`-10\,{\rm Gyr}$).
According to their Equation 3, the observed radii of the Kepler-51 planets can be explained 
if they have about $10\%$ (Kepler-51b), $30\%$ (Kepler-51c), and $40\%$ (KOI-620.02) of their masses in their
H/He envelopes, for the age of $0.3\,{\rm Gyr}$.
Note that the required H/He fractions increase with the system age, because the older planets tend to have smaller radii for fixed masses 
due to the gradual cooling of the gas envelopes.
This may imply that the host star Kepler-51 is actually young, as suggested by the KIC classification (Table \ref{par_star}).
This idea is also supported by \citet{2013MNRAS.436.1883W}, who determined the age of Kepler-51 as $0.53\,\mathrm{Gyr}$,
though they note that the ages are highly uncertain for very young stars.

On the other hand, it seems more difficult to explain how they acquired the above fractions of H/He envelopes.
Although the simulations by \citet{2011ApJ...738...59R} (see, e.g., their Table 2) show that such planets could be formed
by core-nucleated accretion beyond the snow line followed by the inward migration to $T_{\rm eq} \sim 500\,{\rm K}$,
their results are based on the somewhat arbitrary assumption that the planet migrates after the sufficient growth
of its core and envelope.
In situ accretion \citep{2012ApJ...753...66I} is also unlikely to account for the predicted atmospheric fractions, 
unless their natal disk was relatively cool and dissipated slowly.

\begin{table}
	\begin{center}
	\caption{Planet properties obtained from transit light curves and TTVs}
	\label{planet}
	\begin{tabular}{cccc}
	\tableline \tableline
	 Parameter				&		Kepler-51b					&	Kepler-51c					&	KOI-620.02\\
	\tableline													
	$M_p$ ($M_{\oplus}$)		& $2.1_{-0.8}^{+1.5}$			    		& $4.0 \pm 0.4$					& $7.6 \pm 1.1$\\	
	$R_p$ ($R_{\oplus}$)		& $7.1 \pm 0.3$						& $9.0_{-1.7}^{+2.8}$				& $9.7 \pm 0.5$\\	
	$\rho_p$ (${\rm g\,cm^{-3}}$)& $0.03_{-0.01}^{+0.02}$				& $0.03_{-0.03}^{+0.02}$			& $0.046 \pm 0.009$\\
	$a$ (AU) 					& $0.2514 \pm 0.0097$					& $0.384 \pm 0.015$				& $0.509 \pm 0.020$\\	
	$e$ 						& $0.04 \pm 0.01$						& $0.014_{-0.009}^{+0.013}$			& $0.008_{-0.008}^{+0.011}$\\	
	$T_{\rm eq}$ (K)			& $543 \pm 11$						& $439 \pm 9$					& $381 \pm 8$\\			
	\tableline
	\end{tabular}
	\tablecomments{We adopt $M_{\star} = 1.04 \pm 0.12\,M_{\odot}$ in calculating the values of semi-major axes.
	Equilibrium temperatures are calculated from $T_{\rm eff}$ in Table \ref{par_star}, $a/R_{\star}$ in Table \ref{ppar_5th}, and $e$ 
	using $T_{\rm eq} = \sqrt{R_{\star}/2a} (1-e^2)^{-1/4}\,T_{\rm eff}$.}
	\end{center}
\end{table}

In the above analysis, we adopt the value of $M_{\star}$ in Table \ref{par_star}, but not that of $R_{\star}$,
which is only poorly constrained.
More precise determination of the stellar (and hence planetary) mass and radius requires the constraints 
on the stellar parameters with spectroscopic observations.
\section{Analysis of the Double-transit Light Curve: Anomaly Similar to a PPE Event} \label{sec:PPE}
In the double transit of Kepler-51b and KOI-620.02 that occurred around ${\rm BJD} = 2456346.8$,\footnote{
This transit corresponds to the transit number $30$ of Kepler-51b, and number $10$ of KOI-620.02.
Both of these transits were not used in the TTV analysis in Section \ref{sec:TTV}, and so they are not listed in Tables \ref{tc_01} and \ref{tc_02}.}
we identify an increase of the relative flux near the transit center.
Considering the fact that KOI-620 shows $\sim 12\,\mathrm{mmag}$ $(\sim1\%)$ variation associated with its rotation \citep{2013ApJ...775L..11M},
this ``bump" can be naturally explained by a spot-crossing event \citep[e.g.,][]{2003ApJ...585L.147S, 2008ApJ...683L.179S, 2009A&A...494..391R}.
Indeed, we find several transits of KOI-620.02 showing brief brightenings of similar amplitudes ($\sim 0.2\,\%$) 
as seen in this double-transit light curve.
In addition, this double transit occurred during a gradual increase of the stellar flux, which indicates that a large star spot (or a group of star spots)
was moving on the visible side of the star at that time.
However, as we mentioned in the last part of Section \ref{sec:transit}, 
the small difference of their impact parameters obtained from the transit light curves
requires that the PPE {\it should} have occurred in this double transit, provided that
(i) their orbital planes are nearly aligned and that
(ii) their cosine inclinations have the same signs.
Since the inner planet Kepler-51b overtakes the outer one KOI-620.02 in this double transit, 
the minimum sky-plane separation becomes small enough under these conditions.

\begin{figure}
	\begin{center}
	\includegraphics[width=8.5cm,clip]{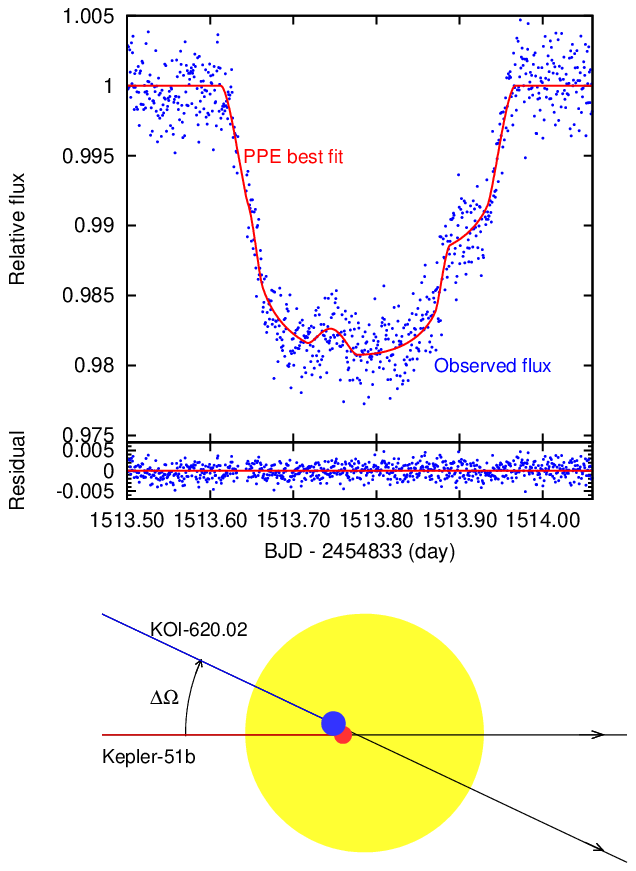}
	\caption{
	Upper: best-fit PPE model (red solid line) for the observed double-transit light curve (blue points).
	Errors of the observed fluxes are omitted for clarity.
	Lower: trajectories of the two planets for the best-fit PPE model. This is a snapshot at the time when the two planets 
	are closest in the plane of the sky.
	An animation of this model is also available at \protect\url{http://www-utap.phys.s.u-tokyo.ac.jp/\~{}masuda/ppe\_animation.gif}.}
	\label{PPE_best}
	\end{center}
\end{figure}

Motivated by this fact, we fit the observed double-transit light curve with the PPE model by \citet{2013ApJ...778..185M}
for $\Delta \Omega$, the longitude of the ascending node of KOI-620.02 relative to that of Kepler-51b.
Note that here we choose the plane of the sky as a reference plane, and so $\Delta \Omega$ corresponds to the mutual inclination 
of the two planetary orbits in this plane (see also the lower panel of Figure \ref{PPE_best}).
Also note that, in the following, we consider a general case where $\Delta \Omega$ can take any value from $-180^{\circ}$ to $180^{\circ}$,
though we only discussed the aligned ($\Delta \Omega \sim 0$) case above as a motivation for the PPE scenario.
This is because, in general, the occurrence of a PPE event is not limited to the aligned case \citep{2013ApJ...778..185M}.
The other parameters $R_p/R_{\star}$, $b$, and $\rho_{\star}$ are also floated
except for $u_1$ and $u_2$, which are fixed at the values in Table \ref{ppar_5th}.
While we restrict the impact parameter $b$ of Kepler-51b to be positive, we allow $b$ of KOI-620.02 to be either positive or negative,
taking into account that the two planets can have different signs of cosine inclinations.
Using an MCMC algorithm, we find that this model gives a reasonably good fit with $\chi^2/{\rm d.o.f} = 0.94$, and obtain
$\Delta \Omega = -25.3_{-6.8}^{+6.2}\,{\rm deg}$ for the sky-plane
mutual inclination of the two planets (Figure \ref{PPE_best} and Table \ref{PPE_fit}).

This value, if true,
indicates that the orbital planes of Kepler-51b and KOI-620.02 are significantly misaligned, 
which means that either of their orbital axes are tilted with respect to the stellar spin axis.
This result may be in contrast to the spin-orbit alignments observed in five multi-transiting systems so far 
\citep{2012Natur.487..449S, 2012ApJ...756...66H, 2012ApJ...759L..36H,  2013ApJ...766..101C, 2013ApJ...771...11A},
but agrees with the recent discovery that the spin-orbit misalignment is not confined to hot-Jupiter systems \citep{2013Sci...342..331H}.

However, if their orbits are really significantly misaligned, it follows that this multi-transiting system is a rare object.
For example, the conditional probability $p(02|{\rm b})$ that the outer KOI-620.02 transits when the inner Kepler-51b is known to transit is 
$p(02|{\rm b}) \simeq a_{02} \sin \phi / R_\star \sim 1/60$ for the mutual inclination of $\phi \sim 30\,{\rm deg}$ and $a_{02}/R_\star \sim 130$ 
\citep[see, e.g.,][]{2010arXiv1006.3727R}.
This value is smaller by the factor of $a_{\rm b} \sin \phi / R_{\star} \sim 30$ than in the aligned case,
where $p(02|{\rm b}) \simeq a_{\rm b}/a_{02} \sim 1/2$.

\begin{table}[h]
	\begin{center}
	\caption{Resulting parameters of the PPE fit to the double-transit light curve}
	\label{PPE_fit}
	\begin{tabular}{c@{\hspace{2cm}}c}
	\tableline \tableline
	Parameter				&		Value\\
	\tableline
	\multicolumn{2}{c}{Kepler-51b}\\
	\tableline
	$a/R_{\star}$				& $63.65 \pm 0.33$\\
	$R_p/R_{\star}$			& $0.0741 \pm 0.0017$\\
	$b$						& $0.016_{-0.012}^{+0.024}$\\
	$t_c ({\rm BJD} - 2454833)$	& $1513.76694 \pm 0.00083$\\
	\tableline
	\multicolumn{2}{c}{KOI-620.02}\\
	\tableline
	$a/R_{\star}$				& $128.93 \pm 0.66$\\
	$R_p/R_{\star}$			& $0.1019 \pm 0.0011$\\
	$b$ 						& $0.039_{-0.040}^{+0.038}$\\
	$t_c ({\rm BJD} - 2454833)$	& $1513.78988 \pm 0.00070$\\
	\tableline
	$\rho_{\star}$				& $2.393 \pm 0.037$\\
	$\Delta \Omega$ 			& $-25.3_{-6.8}^{+6.2}$\\
	$\chi^2/{\rm d.o.f}$		& $807/859$\\
	\tableline
	\end{tabular}
	\tablecomments{In this fit, the impact parameter of KOI-620.02 is allowed to be either positive or negative,
	while that of Kepler-51b is fixed to be positive.}
	\end{center}
\end{table}
\begin{figure}
	\begin{center}
	\includegraphics[width=9cm,clip]{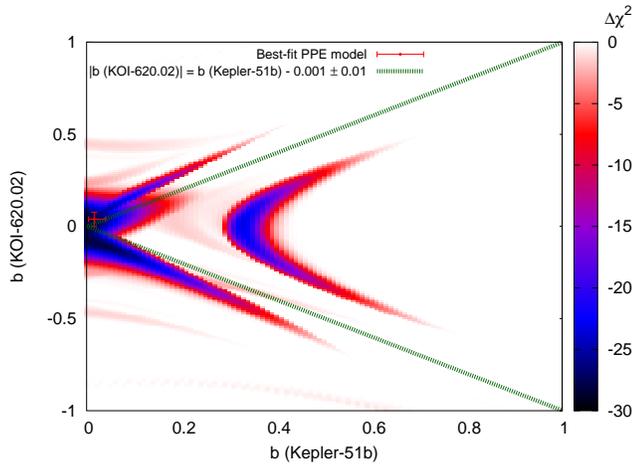}
	\caption{
	Plot to show the region of $b ({\rm Kepler\mathchar`-51b})$-$b ({\rm KOI\mathchar`-620.02})$ plane
	where the PPE model is consistent with the phase-curve analysis. 
	Color scale shows the maximum decrease in $\chi^2$ by including the PPE occurrence into the model;
	each value is calculated for the grid of the impact parameters at the spacing of $0.01$, 
	by varying $\Delta \Omega$ from $-180^{\circ}$ to $180^{\circ}$ at the spacing of $1^{\circ}$.
	The other transit parameters are fixed at the values in Table \ref{PPE_fit}.
	Dotted green lines correspond to the constraint on the impact parameters of the two planets obtained from the phase-folded
	transit light curves in Section \ref{sec:transit}, and a red point with error bars denotes the best-fit values of $b$ in Table \ref{PPE_fit}.
	Note that only the region where $b ({\rm Kepler\mathchar`-51b}) > 0$ is shown in this figure, because the results are symmetric 
	with respect to $(0, 0)$. 
	}
	\label{ppe_search}
	\end{center}
\end{figure}

In fact, a further examination of the PPE model reveals that it is consistent with the results of the phase-curve analysis in Section \ref{sec:transit}
only when both planets have $|b| \sim 0$ as in Table \ref{PPE_fit}, making this scenario all the less likely.
This situation is illustrated in Figure \ref{ppe_search}, where we examine the validity of the PPE model for all the possible values of 
the two planets' impact parameters.
In this plot, the color scale shows the minimum value of the $\chi^2$ difference between the PPE and non-PPE models,
found by incrementing $\Delta \Omega$ by $1^{\circ}$ from $-180^{\circ}$ to $180^{\circ}$.
The grid scale of $b$ is $0.01$, and the other parameters are fixed at the best-fit values in Table \ref{PPE_fit}.
In the dark-blue region, the PPE model significantly improves the fit, 
while the dotted green lines correspond to the constraint from the phase-folded transits we mentioned in Section \ref{sec:transit}.
Comparing these two regions, we find that only the values of $b$ close to the PPE best fit (red point with error bars) 
are consistent with the PPE interpretation of the bump.
Moreover, the impact parameters required by the PPE model are only marginally consistent with those obtained in Section \ref{sec:transit} (Table \ref{ppar_5th}),
though they are close to the ones obtained by the {\it Kepler} team (Table \ref{ppar_prior}).

It should also be noted that, 
if the orbit of the outermost planet has such a large mutual inclination with respect to the inner two,
their orbits will precess rapidly.
In this case, the resulting transit duration variations (TDVs) would be fairly large for the middle grazing planet, Kepler-51c.
Nevertheless, no significant TDVs are apparently seen in the transits of this planet (Figure \ref{phase}).
With no constraints on the nodal angle of Kepler-51c, it may still be possible that the orbit of this planet is also tilted with respect
to that of Kepler-51b so that the effect of precession is canceled out, but it would require fine tuning of the parameters.
Furthermore, we must happen to observe this system when all the planets have small eccentricities 
as indicated from TTVs, while their eccentricities would vary as the system evolves under such a large mutual inclination.

These arguments about the transit probability, impact parameters, and expected orbital precession, 
imply that a PPE event that is consistent with all of the observations would be exceedingly rare;
in other words, the PPE interpretation for the observed anomaly is essentially refuted.
The observed bump is, therefore, probably due to a star spot or just a correlated noise.
If it is the spot crossing, the detailed star-spot modeling may provide valuable information on the stellar obliquity
\citep{2009ApJ...701..756D,  2010A&A...510A..25S, 2011ApJ...733..127S, 2011ApJ...740L..10N, 2012Natur.487..449S, 2013ApJ...775...54S},
which is closely related to the orbital evolution history of the planets \citep[e.g.,][]{2000A&A...359L..13Q, 2005ApJ...631.1215W}.
This is beyond the scope of this paper.

There are several possible approaches to strengthen the above interpretation of the anomaly.
First, follow-up observation of the next 
double transit where a PPE event might occur
is unreasonable, because we have to wait at least until 2092
even if the orbital planes of the two planets are completely aligned.
Secondly, the more accurate determination of the impact parameters would be helpful.
If it is confirmed that the impact parameters of the two planets differ from zero with better precision,
the discussion based on Figure \ref{ppe_search} is enough to exclude the PPE scenario.
In contrast, if both planets have $|b| \sim 0$ as suggested by the {\it Kepler} team (Table \ref{ppar_prior}), 
we need to explain why the two planets {\it did not} overlap;
if $\Delta \Omega \sim 0$ in this case, we should have observed a large bump that is totally inconsistent with the observed one.
In order to better determine the impact parameters, the better constraint on $R_{\star}$ (or $\rho_{\star}$ itself) would again be quite beneficial,
because the prior knowledge on $\rho_{\star}$ pins down their values, which are strongly correlated to that of $\rho_{\star}$.
Lastly, as mentioned above, a thorough analysis of the dynamical model taking account of mutual orbital inclinations and (if necessary) star spots
would provide a more decisive conclusion on the origin of this anomaly.
Yet another possibility is the analysis of the long-term dynamical stability, which may rule out the misaligned configuration.
\section{Summary and Discussion} \label{sec:summary}
We have discussed the two topics in this paper,
characterization of the multi-transiting planetary system around Kepler-51 with TTV analysis (Sections \ref{sec:transit} and \ref{sec:TTV})
and interpretation of the light-curve feature similar to a PPE caused by the two planets in this system (Section \ref{sec:PPE}).
Here we briefly summarize each of the topics and give some additional comments.\\

\noindent \textbf{1. Characterization of the Kepler-51 system.}

We analyzed the transit light curves and TTVs of the three planets in the Kepler-51 system,
which lie close to a $1:2:3$ resonance chain.
Combining the planetary masses obtained from TTVs, and planet-to-star radius ratios and stellar density inferred from the transit light curves,
we determined the properties of the three planets as follows:
$M_{\rm b} = 2.1_{-0.8}^{+1.5} \,M_{\oplus}, R_{\rm b} = 7.1 \pm 0.3 \,R_{\oplus}, \rho_{\rm b} = 0.03_{-0.01}^{+0.02} \,{\rm g\,cm^{-3}}$ for Kepler-51b (KOI-620.01),
$M_{\rm c} = 4.0 \pm 0.4 \,M_{\oplus}, R_{\rm c} = 9.0_{-1.7}^{+2.8} \,R_{\oplus}, \rho_{\rm c} = 0.03_{-0.03}^{+0.02}\,{\rm g\,cm^{-3}}$ for Kepler-51c (KOI-620.03), and
$M_{02} = 7.6 \pm 1.1 \,M_{\oplus}, R_{02} = 9.7 \pm 0.5 \,R_{\oplus}, \rho_{02} = 0.046 \pm 0.009 \,{\rm g\,cm^{-3}}$ for KOI-620.02.
From these results, as well as the sinusoidal modulation consistent with their proximities to the resonances, 
we confirmed KOI-620.02 as a planet in this system (Kepler-51d), which has an equilibrium temperature close to
the inner edge of the habitable zone.

The even more remarkable implication of our analysis is that the densities of all three planets in this system are among the lowest yet determined,
though a more detailed study taking account of the presence of star spots might increase these values.
In fact, such low-density planets are frequently seen in other compact multi-transiting planetary systems;
these include the systems around
Kepler-9 \citep{2010Sci...330...51H}, 
Kepler-11 \citep{2011Natur.470...53L, 2012MNRAS.427..770M, 2013ApJ...770..131L}, 
Kepler-18 \citep{2011ApJS..197....7C}, 
Kepler-30 \citep{2012Natur.487..449S}, 
Kepler-56 \citep{2013Sci...342..331H}, 
Kepler-87 \citep{2014A&A...561A.103O}, 
Kepler-89 \citep{2013ApJ...768...14W, 2013ApJ...778..185M}, and 
KOI-152 \citep{2013arXiv1310.2642J}, 
all of which have planets with sub-Saturn densities.
Considering the fact that such systems frequently contain giant planets and 
planet pairs in near mean-motion resonances (and, of course, that they are the multi-transiting systems), 
they were probably formed via the convergent disk migration and subsequent resonance capture. 
As pointed out by \citet{2013arXiv1310.2642J}, the selection effects for planets suitable for the TTV analysis could be significant.
Nevertheless, if the low-densities observed so far are intrinsic features of compact multi-transiting systems,
they could be an important constraint on the formation mechanisms via disk migrations.
\\

\noindent \textbf{2. Anomaly similar to a planet-planet eclipse event.}

We also analyzed the double-transit light curve of Kepler-51b and KOI-620.02 around ${\rm BJD} = 2456346.8$.
The archived {\it Kepler} light curve shows a slight increase in the relative flux of Kepler-51,
which could be explained by the PPE (planet-planet eclipse), the overlap of the two planets during their double-transit phase.

If the cosine inclinations of the two planets have the same signs, the impact parameters of the two planets
strongly suggest that the PPE should have occurred in this double transit.
Indeed, the PPE model well reproduces the observed anomaly for the sky-plane mutual inclination between the two planets of $\sim 25\,{\rm deg}$,
which implies that their orbital planes are misaligned.
This result, if true, indicates that either of their orbital planes are tilted with respect to the stellar spin axis, and
makes the Kepler-51 system another important piece of evidence that the spin-orbit misalignment is not confined to
hot-Jupiter systems \citep{2013Sci...342..331H}.

However, this interpretation of the anomaly seems unlikely for the following reasons.
First, such a large mutual inclination significantly reduces the probability that both of Kepler-51b and KOI-620.02 transit.
Second, the PPE model is consistent with the result of phase-curve analysis only for limited values of the two planets' impact parameters.
Finally, the misaligned configuration would result in the rapid orbital precession, whose effect should have been readily detectable
in the transit light curves of the middle grazing planet, Kepler-51c.
Alternative interpretations of the anomaly include the correlated noise and the star-spot crossing.
If the latter is the case, it may provide us the information on the stellar obliquity 
\citep{2009ApJ...701..756D, 2010A&A...510A..25S, 2011ApJ...733..127S, 2011ApJ...740L..10N, 2012Natur.487..449S, 2013ApJ...775...54S},
which is definitely valuable in unveiling the orbital evolution history of the planets in this system.

In any case, it is rewarding to explore the origin of this anomaly, because it serves as an example of the false positive of a PPE event.
Compared with the case of the Kepler-89 (KOI-94) system, where a small light-curve modulation led to the clear detection of a PPE \citep{2012ApJ...759L..36H},
the situation is less ideal for the Kepler-51 system analyzed in this paper.
A detailed investigation of the possible phenomena (e.g., star spots) that could produce PPE-like features
would help the future detection of this valuable event in such marginal conditions.
\acknowledgments 
This work is based on the photometry of Kepler-51 provided by the {\it Kepler} mission,
and the author gratefully acknowledges the {\it Kepler} team.
The author thanks Masahiro Ikoma for helpful discussion on the interpretation of the observed planetary densities.
The author would also like to thank an anonymous referee, Dan Fabrycky,  Jack Lissauer, Sean Mills, and
Yasushi Suto for useful comments to improve the manuscript.
This work was supported by the Program for Leading Graduate Schools, MEXT, Japan.\\



\end{document}